\pdfoutput=1
\documentclass[twocolumn]{aastex62}

\usepackage{amsmath}

\received{9th June 2020}
\accepted{16th September 2020}

\submitjournal{ApJ}
\begin{document}

\title{WASP-117 b: an eccentric hot-Saturn as a future complex chemistry laboratory}

\author[0000-0002-7771-6432]{Lara O. Anisman}
\affil{Department of Physics and Astronomy, University College London, London, United Kingdom}

\author[0000-0002-5494-3237]{Billy Edwards}
\affil{Department of Physics and Astronomy, University College London, London, United Kingdom}

\author[0000-0001-6516-4493]{Quentin Changeat}
\affil{Department of Physics and Astronomy, University College London, London, United Kingdom}

\author[0000-0003-2854-765X]{Olivia Venot}
\affil{Laboratoire Interuniversitaire des Systèmes Atmosphériques (LISA), UMR CNRS 7583, Université Paris-Est-Créteil, Université de Paris, Institut Pierre Simon
Laplace, Créteil, France}

\author[0000-0003-2241-5330]{Ahmed F. Al-Refaie}
\affil{Department of Physics and Astronomy, University College London, London, United Kingdom}

\author[0000-0003-3840-1793]{Angelos Tsiaras}
\affil{Department of Physics and Astronomy, University College London, London, United Kingdom}

\author[0000-0001-6058-6654]{Giovanna Tinetti}
\affil{Department of Physics and Astronomy, University College London, London, United Kingdom}

\begin{abstract}

   We present spectral analysis of the transiting Saturn-mass planet WASP-117\,b, observed with the G141 grism of the Hubble Space Telescope’s Wide Field Camera 3 (WFC3).  We reduce and fit the extracted spectrum from the raw transmission data using the open-source software Iraclis before performing a fully Bayesian retrieval using the publicly available analysis suite TauREx 3.0. We detect water vapour alongside a layer of fully opaque cloud, retrieving a terminator temperature of $T_{term}=833^{+260}_{-156}$ K. In order to quantify the statistical significance of this detection, we employ the Atmospheric Detectability Index (ADI), deriving a value of ADI = 2.30, which provides positive but not strong evidence against the flat-line model. Due to the eccentric orbit of WASP-117\,b, it is likely that chemical and mixing timescales oscillate throughout orbit due to the changing temperature, possibly allowing warmer chemistry to remain visible as the planet begins transit, despite the proximity of its point of ingress to apastron. We present simulated spectra of the planet as would be observed by the future space missions Ariel and JWST and show that, despite not being able to probe such chemistry with current HST data, these observatories should make it possible in the not too distant future. \vspace{15mm}
\end{abstract}

\section{Introduction}

Among the gaseous exoplanets detected so far, a small subset  are Saturn-mass ($\sim0.3 M_J$) with radii larger than $1R_{J}$, making their atmospheres inflated. Examples include WASP-69\,b, Kepler-427\,b, WASP-151\,b and HAT-P-51\,b \citep{Anderson_2014,H_brard_2014,Demangeon_2018,Hartman_2015}. Most of these inflated hot-Saturns have been discovered using the transit method, which provides an observational bias towards lower eccentricity systems. Thus, eccentric hot-Saturns that have been discovered so far are fairly limited, including GJ\,1148\,b, which is not transiting and for which the radius is not constrained \citep{Trifonov_2018}, and HAT-P-19\,b, which is indeed inflated but has a relatively low eccentricity of $e\approx0.067$ \citep{Hartman_2010}. At the time of its discovery in 2014, WASP-117\,b was the first planet found to possess a period larger than 10 days by the WASP survey, and at present remains the lowest mass gaseous planet with such a period, with a mass of $M_{p} = 0.2755 \pm 0.0089$ $M_{J}$ and a radius of $R_{p} = 1.021^{+0.076}_{-0.065}$ $R_{J}$ \citep{lendl_wasp117}. With a well-constrained eccentricity of $e = 0.302 \pm 0.023$,  WASP-117\,b exhibits itself as an inflated Saturn-mass planet in an eccentric, misaligned orbit around a bright (V$_{mag}$ = 10.15) main-sequence F9 star, a rarity among transiting gaseous extra-solar planets.
\\

As a consequence of its large orbital distance, the tidal forces exerted on WASP-117\,b by its host star are thought to be weak, making the planet’s eccentric orbit very stable over the system lifetime. Subsequently this planet provides an important case study for analysis of orbital dynamics and disc migration in gaseous exoplanets, as alluded to in \citet{lendl_wasp117}. The eccentric nature of its orbit gives rise to fluctuations in the stellar flux received by the planet. This results in a variation in the temperature as WASP-117\,b traverses its orbit, which in turn may cause changes in the corresponding chemistry. The thermal variations caused by the orbital parameters coupled with large chemical mixing timescales make this planet a tantalising and, at present, unique object for the study of exoplanet atmospheric chemistry. Fortunately, the lack of significant activity and brightness of its host star \citep{lendl_wasp117} make WASP-117\,b an excellent candidate for atmospheric characterisation.\\

In recent years the Hubble and Spitzer Space Telescopes have enabled the study of an increasing number of exoplanetary atmospheres through transit, eclipse, or phase-curve spectro-photometric observations \citep[e.g.][]{vidal,swain2008molecular,laughlin,Linsky_2010,Tinetti_2010,Majeau_2012,Deming_2013,Fraine_2014,Stevenson_2014,Morello_2016,Evans_2017,skaf_ares,edwards2020ares}.
Complementary observations from the ground, through high-dispersion or direct imaging spectroscopic techniques, have allowed for the extension of atmospheric observations to non-transiting planets \citep[e.g.][]{Macintosh_2015,Brogi_2012}.  
While a handful of smaller planets have been observed \citep[e.g.][]{Kreidberg_GJ1214b_clouds,tsiaras_55cnce,de_Wit_2018,Demory_2016,Kreidberg_2019,tsiaras_h2o,benneke_k2-18}), the current sample of observed exoplanetary atmospheres is still biased towards larger planets, which typically present a stronger signal to detect \citep[e.g.][]{sing,iyer_pop,tsiaras_30planets,pinhas}. \\

In this paper we present an analysis of the HST WFC3 G141 transmission spectrum of WASP-117\,b. Our retrievals show evidence of water vapour but, due to the narrow spectral coverage (1.088 - 1.688 $\mu m$), we are unable to constrain the abundances of carbon-based molecules such as CH$_4$, CO and CO$_2$. The molecule which is perhaps most indicative of potential chemical changes over the orbit of WASP-117\,b due to orbit-induced temperature variations is CH$_4$. Future space observatories and missions like JWST \citep{Greene_2016} and Ariel \citep{tinetti_ariel}, with long spectral baselines, will enable us to widen and deepen our spectral view and subsequently reveal possible complex chemistry. We present simulations of equilibrium chemical profiles at WASP-117\,b's temperature extremes to demonstrate that, while the HST data is insufficient to distinguish between these cases, Ariel and JWST should have the precision and spectral coverage to disentangle these scenarios.

\vspace{18mm}

\section{Methods}
\subsection{HST-WFC3 Data Analysis}

Our analysis of the HST-WFC3 data started from the raw spatially scanned spectroscopic images which were obtained from the Mikulski Archive for Space Telescopes\footnote{\url{https://archive.stsci.edu/hst/}}. The transmission spectrum of WASP-117 b was acquired by proposal 15301 and was taken in September 2019. We used Iraclis\footnote{\url{https://github.com/ucl-exoplanets/Iraclis}}, a specialised, open-source software for the analysis of WFC3 scanning observations. The reduction process included the following steps: zero-read subtraction, reference pixels correction, non-linearity correction, dark current subtraction, gain conversion, sky background subtraction, calibration, flat-field correction, and corrections for bad pixels and cosmic rays. For a detailed description of these steps, we refer the reader to the original Iraclis papers \citep{tsiaras_hd209, tsiaras_30planets, tsiaras_55cnce}.\\

\begin{figure}
    \centering
    \includegraphics[width = \columnwidth]{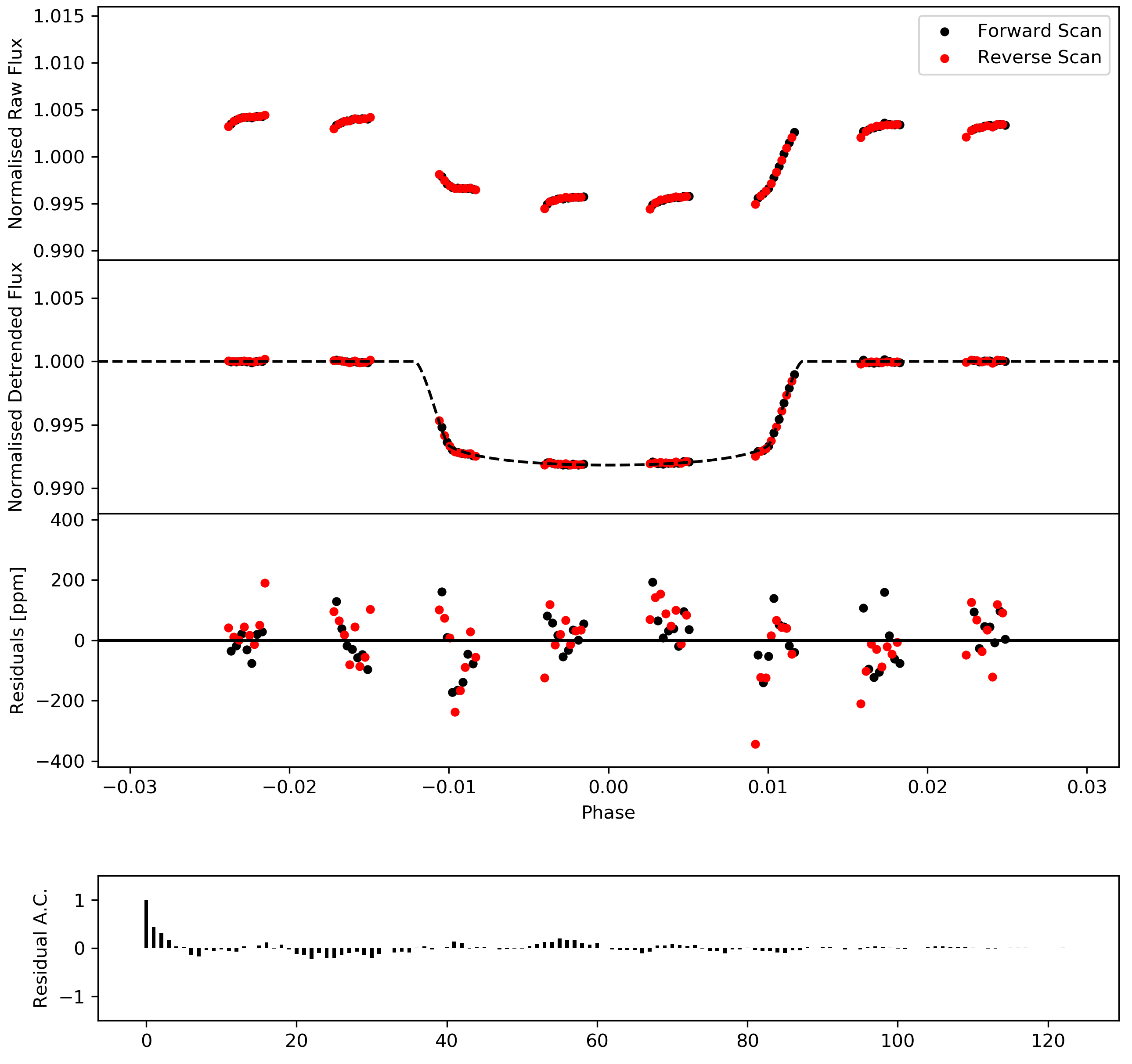}
    \caption{White light curve for the transmission observation of WASP-117 b. First panel: raw light curve, after normalisation. Second panel: light curve, divided by the best fit model for the systematics. Third panel: residuals for best-fit model. Fourth panel: auto-correlation function of the residuals.}
    \label{fig:white}
\end{figure}

\begin{figure}
    \centering
    \includegraphics[width = \columnwidth]{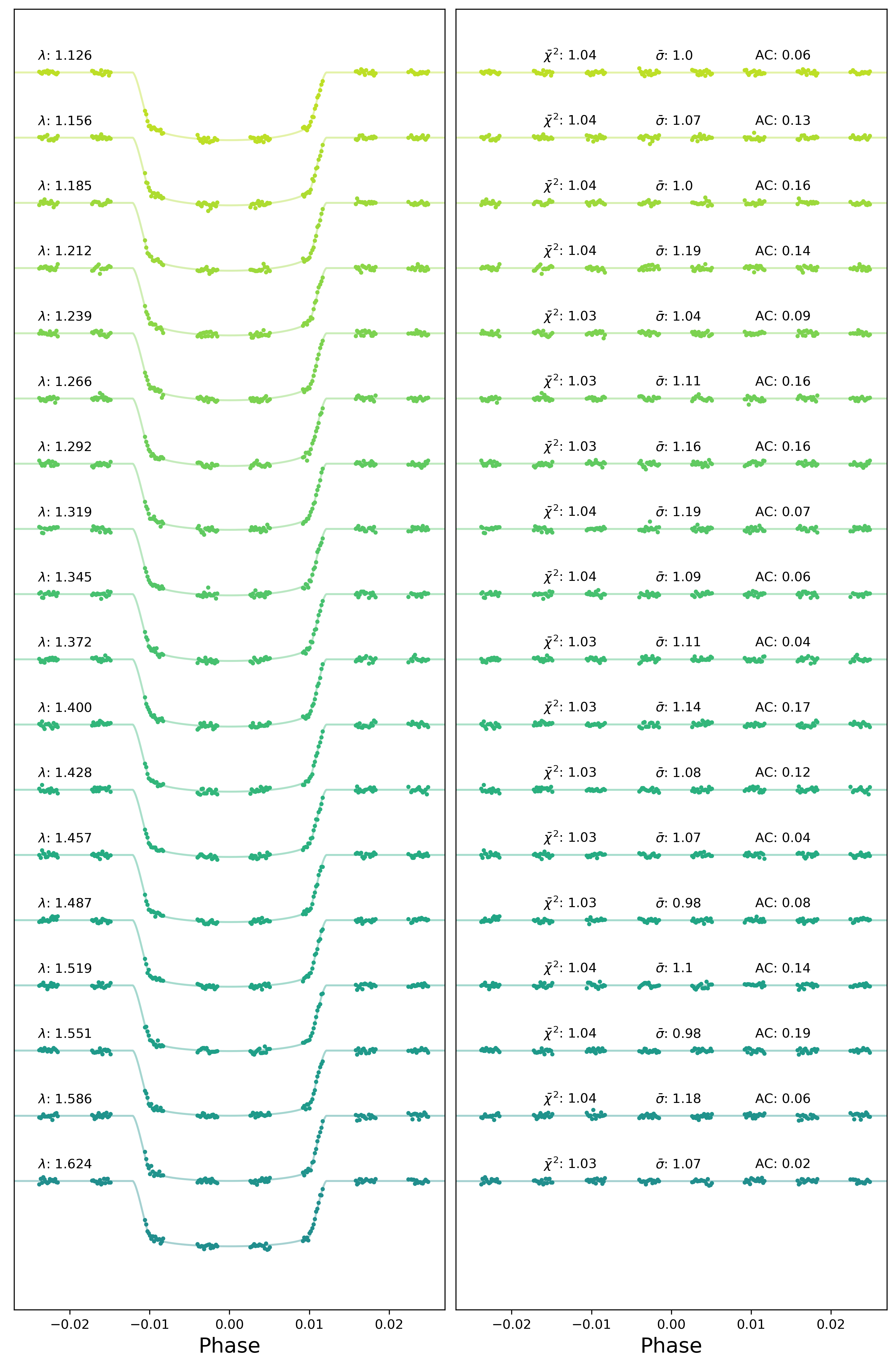}
    \caption{Spectral light curves fitted with Iraclis for the transmission spectra where, for clarity, an offset has been applied. Left: the detrended spectral light curves with best-fit model plotted. Right: residuals from the fitting with values for the Chi-squared ($\chi^2$), the standard deviation of the residuals with respect to the photon noise ($\bar{\sigma}$) and the auto-correlation (AC).}
    \label{fig:spectral}
\end{figure}

The reduced spatially scanned spectroscopic images were then used to extract the white (1.088 - 1.688 $\mu$m) and spectral light curves. As is routinely done for HST studies, we then discarded the first orbit of the visit as it presents stronger wavelength pendant ramps. For the fitting of the white light curve, the only free parameters were the mid-transit time and planet-to-star ratio, with other values fixed to those from \citet{lendl_wasp117} ($P = 10.020607$, $a/R_s = 17.39$, $i = 89.14$, $\omega = 242$, $T_0 = 2457355.51373$). However, the white light curve fit showed significant residuals. We therefore fitted the light curve with the reduced semi-major axis, $a/R_s$, as an additional free parameter. We then performed a final white light curve fitting with our updated value of $a/R_s=17.65$. While some residuals remain, the divide-by-white method ensures these are not seen in the spectral light curves. The limb-darkening coefficients were selected from the best available stellar parameters using values from \citet{claretI,claretII} and using the stellar parameters from \citet{lendl_wasp117}. The fitted white light curve for the transmission observation is shown in Figure \ref{fig:white} while the spectral light curves are plotted in Figure \ref{fig:spectral}.

{}

\subsection{Ephemeris Refinement}

Accurate knowledge of exoplanet transit times is fundamental for atmospheric studies. To ensure that WASP-117\,b can be observed in the future, we used our HST white light curve mid time, along with data from TESS \citep{ricker}, to update the ephemeris of the planet. TESS data is publicly available through the MAST archive and we use the pipeline from \citet{edwards_orbyts} to download, clean and fit the 2 minute cadence Pre-search Data Conditioning (PDC) light curves \citep{smith_pdc,stumpe_pdc1,stumpe_pdc2}. WASP-117\,b had been studied in Sectors 2 and 3 and, after excluding bad data, we recovered 4 transits. These were fitted individually with the planet-to-star radius ratio ($R_p/R_s$), reduced semi-major axis ($a/R_s$), inclination ($i$) and transit mid time ($T_{mid}$) as free parameters. \vspace{3mm}

\subsection{Atmospheric Modelling}

Due to the fact that WASP-117\,b is in possession of an eccentric and misaligned orbit, it is thought that its atmosphere may exhibit significant changes in temperature as it traverses its orbit. We can estimate the temperature range by calculating the equilibrium (day-side) temperature expected at periastron and apastron, which we have calculated to be at a distance of approximately 0.067 AU and 0.124 AU from the host star, respectively. The day-side equilibrium temperature of a planet, $T_{p}$, at a distance $a$ from its host star can be derived as a result of equating the incident stellar flux on the planet with that which is absorbed by the planet:
\begin{equation}
T_{p} =  T_{*}\sqrt{\frac{R_{*}}{2a}}\left(\frac{1 - A}{\beta\epsilon}\right)^{\frac{1}{4}} 
%\label{eq:temp}
\end{equation}

as given in \citet{M_ndez_2017}, where $\beta$ is a measure of the fraction of surface area over which the planet re-radiates the stellar flux that it absorbs, $\epsilon$ is the broadband thermal emissivity, and $A$ is the planetary surface albedo. The temperature and radius of the star are denoted as $T_{*}$ and $R_{*}$, respectively.
\\

Considering the eccentricity of the orbit, it is likely that WASP-117\,b is not tidally locked, possibly allowing for effective heat redistribution. Hence, using equation 1 with these assumptions, we take $\beta=1$ and $A=0.3$, obtaining equilibrium temperatures of $T_{p}=1116$ K and $T_{p}=817$ K at periastron and at apastron, respectively. Since the planet has an argument of periastron determined by \citet{lendl_wasp117} as  $\omega = 242.0^{+2.3}_{-2.7}$ degrees, we obtain a transit equilibrium temperature of $T_{p}=838$ K and expect to probe the terminator region of its atmosphere with the planet very close to its least irradiated region of its orbit.
\begin{figure}
    \centering
    \includegraphics[width=0.98\columnwidth]{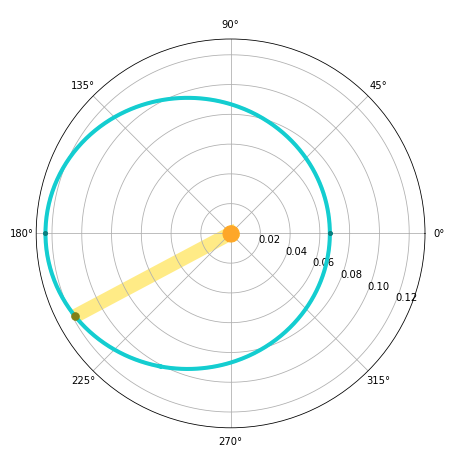}
    \caption{A schematic diagram of the orbital trajectory of WASP-117\,b, where the shaded yellow region illustrates the line-of-sight direction that we observe during transit, confirming the proximity to apastron of this region.}
    \label{fig:orbit}
\end{figure} 
We have estimated the possible chemical composition of the atmosphere by running equilibrium chemistry models for WASP-117\,b at its temperature extremes. More specifically we used the ACE equilibrium chemistry package \citep{Ag_ndez_2012,Venot_2012} contained in TauREx 3.0 to generate  molecular abundances at the periastron and apastron equilibrium temperatures of 1100K and 800K, respectively. 
Together with input stellar parameters, given in Table \ref{tab:star_params}, and retrieved values from WFC3 data for the planet's radius and cloud pressure, these abundance profiles were then used as input for free-chemical forward models at the retrieved terminator temperature, in order to investigate what sort of chemistry might be visible during transit. 
\\

\begin{table}
\centering
\begin{tabular}{cccc}
\hline\hline
\multicolumn{4}{c}{Transmission Spectrum}\\ \hline
Wavelength & Transit Depth & Error & Bandwidth \\
\hline
$[ \mu m]$ & [\%] & [\%] & $[ \mu m]$ \\
\hline
1.12620 & 0.74725 & 0.00412 & 0.03080 \\
1.15625 & 0.74810 & 0.00417 & 0.02930 \\
1.18485 & 0.75177 & 0.00379 & 0.02790\\
1.21225 & 0.74688 & 0.00457 & 0.02690\\
1.23895 & 0.74374 & 0.00425 & 0.02650\\
1.26565 & 0.75076 & 0.00425 & 0.02690\\
1.29245 & 0.74048 & 0.00453 & 0.02670\\
1.31895 & 0.74098 & 0.00449 & 0.02630\\
1.34535 & 0.74622 & 0.00415 & 0.02650\\
1.37230 & 0.75244 & 0.00420 & 0.02740\\
1.4000 & 0.75249 & 0.00434 & 0.02800\\
1.42825 & 0.75465 & 0.00409 & 0.02850\\
1.45720 & 0.75548 & 0.00438 & 0.02940\\
1.48730 & 0.75104 & 0.00379 & 0.03080\\
1.51860 & 0.74379 & 0.00422 & 0.03180\\
1.55135 & 0.73922 & 0.00410 & 0.03370\\
1.58620 & 0.74068 & 0.00464 & 0.03600\\
1.62370 & 0.74473 & 0.00421 & 0.03900\\

\hline \hline
\end{tabular}
\caption{Reduced and fitted spectral data from the raw HST-WFC3 transmission data using Iraclis. \vspace{5mm}}
\label{tab:transmission_data}
\end{table}

\subsection{Spectral Retrieval Simulations}

In order to extract the information content of WASP-117\,b's WFC3 transmission spectrum, retrieval analysis was performed using the publicly available retrieval suite TauREx 3.0 \citep{waldmann_2,waldmann_1,al-refaie_taurex3}\footnote{\url{https://github.com/ucl-exoplanets/TauREx3_public}}, in addition to performing retrievals on our simulated Ariel and JWST spectra, discussed in Section 2.6. For the stellar parameters and the planet mass, we used the values from \cite{lendl_wasp117}, as given in Table \ref{tab:star_params}. In our runs we assumed that WASP-117 b possesses a primary atmosphere with fill gas abundance ratio of $V_{\text{He}}/V_{\text{H}_2}$ = 0.17, where $V_{x}$ denotes the volume mixing ratio for molecule $x.$ We included in our simulations the contribution of trace gases whose opacities were taken from the ExoMol \citep{Tennyson_exomol}), HITRAN \citep{gordon} and HITEMP \citep{rothman} databases for: H$_2$O \citep{polyansky_h2o}, CH$_4$ \citep{exomol_ch4}, CO \citep{li_co_2015} and CO$_2$ \citep{rothman_hitremp_2010}. 
Additionally, we included the Collision Induced Absorption (CIA) from H$_2$-H$_2$ \citep{abel_h2-h2, fletcher_h2-h2} and H$_2$-He \citep{abel_h2-he}, as well as Rayleigh scattering for all molecules. In our retrieval analysis, we used uniform priors for all parameters as described in Table \ref{tab:retrieval_params}. Finally, we explored the parameter space using the nested sampling algorithm MultiNest \citep{Feroz_multinest} with 1500 live points and an evidence tolerance of 0.5.

\vspace{15mm}
\subsection{Atmospheric Detectability}
We quantify the significance of our retrieval results by adopting the formalism of the Atmospheric Detectability Index (ADI) introduced in \citet{tsiaras_30planets}. The ADI is defined as the Bayes Factor, or likelihood ratio, between the retrieved atmospheric model (R) and the flat-line model (F), where the latter is designed to include known degeneracies between parameters in both models. Using the Bayes evidence of each model, as calculated as part of the retrieval, we may determine the ADI as follows:
\begin{equation}
  \text{ADI} =
    \begin{cases}
      \ln\left(\frac{E_{R}}{E_{F}}\right) \quad \text{if} \quad  \frac{E_{R}}{E_{F}} > 1 \\
      0 \quad \quad \quad \quad \text{otherwise}
    \end{cases}       
\end{equation}

where $E_{R}$ and $E_{F}$ are the Bayes evidence for the retrieval model and the flat-line model, respectively. In our case, the retrieval model included the following contributions to opacity: molecular, simple fully-opaque clouds, collision-induced absorption due to H$_2$-H$_2$ and H$_2$-He, along with Rayleigh scattering. As for the flat-line model, we included only simple fully-opaque clouds. This ensures that the derived value of ADI gives a detection significance for the atmosphere detected whilst known degeneracies in each model between the radius of the planet, its temperature and the height of possible clouds, have been accounted for.
\\

\begin{table}
\centering
\begin{tabular}{cccc}
\hline\hline
\multicolumn{2}{c}{Stellar \& Planetary Parameters}\\ 
\hline
    Parameter & Value\\ \hline
    $T_{*}$ [K] & $6038$   \\
    $R_{*}$ [$R_\odot$] & 1.170  \\
    $M_{*}$ [$M_\odot$] & 1.126  \\
    $\log_{10}(g)_{*}$ [$cm/s^{2}$] &4.28 \\
    $[\text{Fe/H}]_{*}$ &  -0.11  \\
    $e$ & $0.302$  \\
    $i$ [deg] & $89.14$  \\
    $\omega$ [deg] & $242.0$ \\ 
    $\Psi$ [deg] & 69.6 \\
    $M_{p}$ [$M_{J}$] & 0.276 \\
    $R_{p}$ [$R_{J}$]& 1.021 \\
    $P_{orbital}$ [days] &  10.02 \\
    \hline \hline
    \end{tabular}
    \caption{Stellar and planetary parameters for WASP-117\,b, for input into Iraclis and TauREx 3.0, derived from \citet{lendl_wasp117}.}
    \label{tab:star_params}
\end{table}

\subsection{Ariel \& JWST simulations}

Following on from Section 2.3, in order to investigate observable chemistry on WASP-117\,b, we have simulated two different spectra assuming the planet and stellar parameters as specified in Table \ref{tab:star_params} and using TauREx 3.0 to generate chemical equilibrium forward models. Both forward models are created using the retrieved terminator temperature of 833 K, with one using chemical equilibrium molecular abundances expected for a 1100 K atmosphere and one using those for one at 800 K. We note that this does not account for disequilibrium processes such as quenched molecular abundances to deep atmospheric levels due to vertical mixing, for example. During its primary mission, Ariel will survey the atmospheres of 1000 exoplanets \citep{edwards_ariel} while JWST could observe up to 150 over the 5-year mission lifetime \citep{cowan}. WASP-117\,b is an excellent target for characterisation with either observatory and so we generate error bars for the simulated spectra using ArielRad \citep{mugnai} and ExoWebb \citep{exowebb}. For JWST we modelled observations with NIRISS GR700XD (0.8 - 2.8 $\mu m$) and NIRSpec G395M (2.9 - 5.3 $\mu m$), assuming 2 transit observations with each instrument whilst for Ariel, which provides simultaneous coverage from 0.5 - 7.8 $\mu m$, we simulated error bars at tier 3 resolution for 15 transit observations.

\section{Results}

\subsection{HST-WFC3 Data Analysis and Interpretation}

\begin{figure}
    \centering
    \includegraphics[width = 9.5cm]{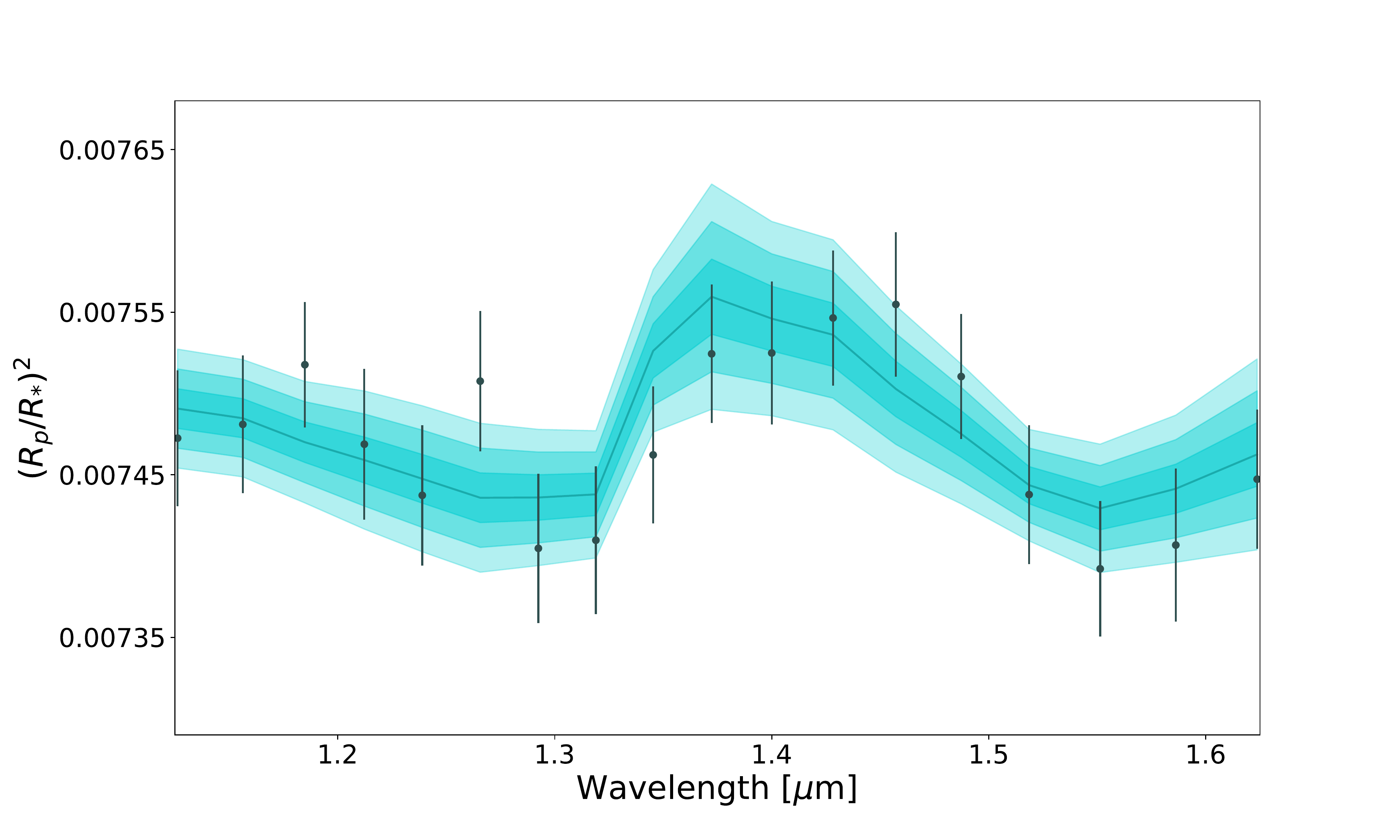}
    \caption{Best-fit transmission spectrum of WASP-117 b.}
    \label{fig:my_label}
\end{figure}

Our retrieval analysis determined the presence of water vapour with a volume mixing ratio, $V_{\text{H}_2\text{O}}$, given by $\log_{10}(V_{\text{H}_{2}\text{O}})= -3.82 ^{+1.37}_{-1.55}$, in the atmosphere of WASP-117\,b.
Additionally a layer of grey cloud at $10^{2.52^{+1.53}_{-1.25}}$ Pa was retrieved, with the parameter $P_{cloud} = \frac{P \text{[Pa]}}{1 \text{[Pa]}}$ denoting the ratio between the atmospheric pressure at which the cloud layer sits, and 1 Pa, to provide a dimensionless argument for the logarithm. The corresponding transmission data and fitted spectrum are displayed in Figure \ref{fig:my_label} and Table \ref{tab:transmission_data}.

\begin{figure*}
    \centering
    \includegraphics[width=0.97\textwidth]{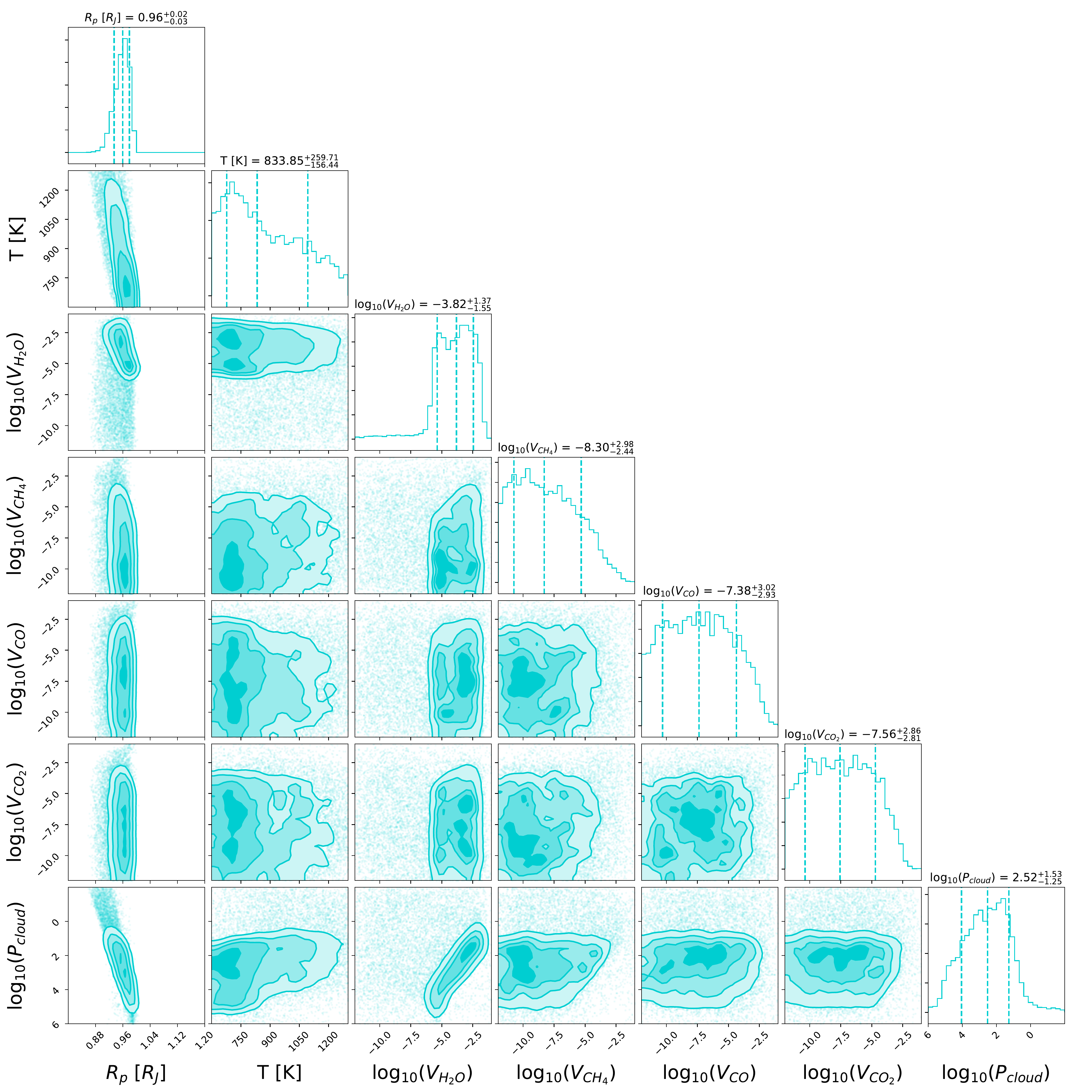}
    \caption{Posterior distributions for the transmission spectrum (see Figure \ref{fig:my_label}) of WASP-117 b which indicate the presence of water vapour and cloud.}
    \label{fig:post}
\end{figure*} 

The abundance of water retrieved is consistent with results from population studies of gaseous planets such as \citet{tsiaras_30planets}, \citet{pinhas} and \citet{sing} and chemistry models of gaseous atmospheres (e.g. \cite{Venot_2012}). While we did attempt to retrieve other trace gases such as CO, CH$_4$ and CO$_2$, we were unable to identify their presence. Our detected abundance of water vapor is consistent with a similar study of WASP-117\,b, \citet{carone_w117}, to within 1$\sigma$, however we find that the data do not constrain said abundance to great accuracy ($\log_{10}(V_{\text{H}_{2}\text{O}})$ \text{at} $1\sigma \in$ [-5.37, 2.45]). In addition we do not find evidence for the presence of other species, thus we are not able to adequately constrain atmospheric metallicity for this planet. The priors used in our retrieval run as well as the retrieved values are summarised in Table \ref{tab:retrieval_params}. The full posterior distribution for the parameters is shown in Figure \ref{fig:post}.
In the case of WASP-117\,b, we detect our atmospheric retrieval signature at the ADI value of 2.30 which as in \citet{Kass1995bayes} corresponds to positive evidence against the flatline model.

\subsection{Ephemeris Refinement}
The transits of WASP-117\,b from HST and TESS were seen to arrive early compared to the predictions from \citet{lendl_wasp117}. The ephemeris of WASP-117\,b was recently refined by \citet{mallonn}. We used the observations from \citet{mallonn}, the original ephemeris from \citet{lendl_wasp117}, and the new data analysed here to update the period and transit time for the planet. Using this data, we determined the ephemeris of WASP-117\,b to be $P = 10.0205928 \pm 0.0000044$ days and $T_0$ = 2458688.251803 $\pm$ 0.000097 BJD$_{TDB}$ where $P$ is the planet's period, $T_0$ is the reference mid-time of the transit and BJD$_{TDB}$ is the barycentric Julian date in the barycentric dynamical.

\begin{figure}
    \centering
    \includegraphics[width = \columnwidth]{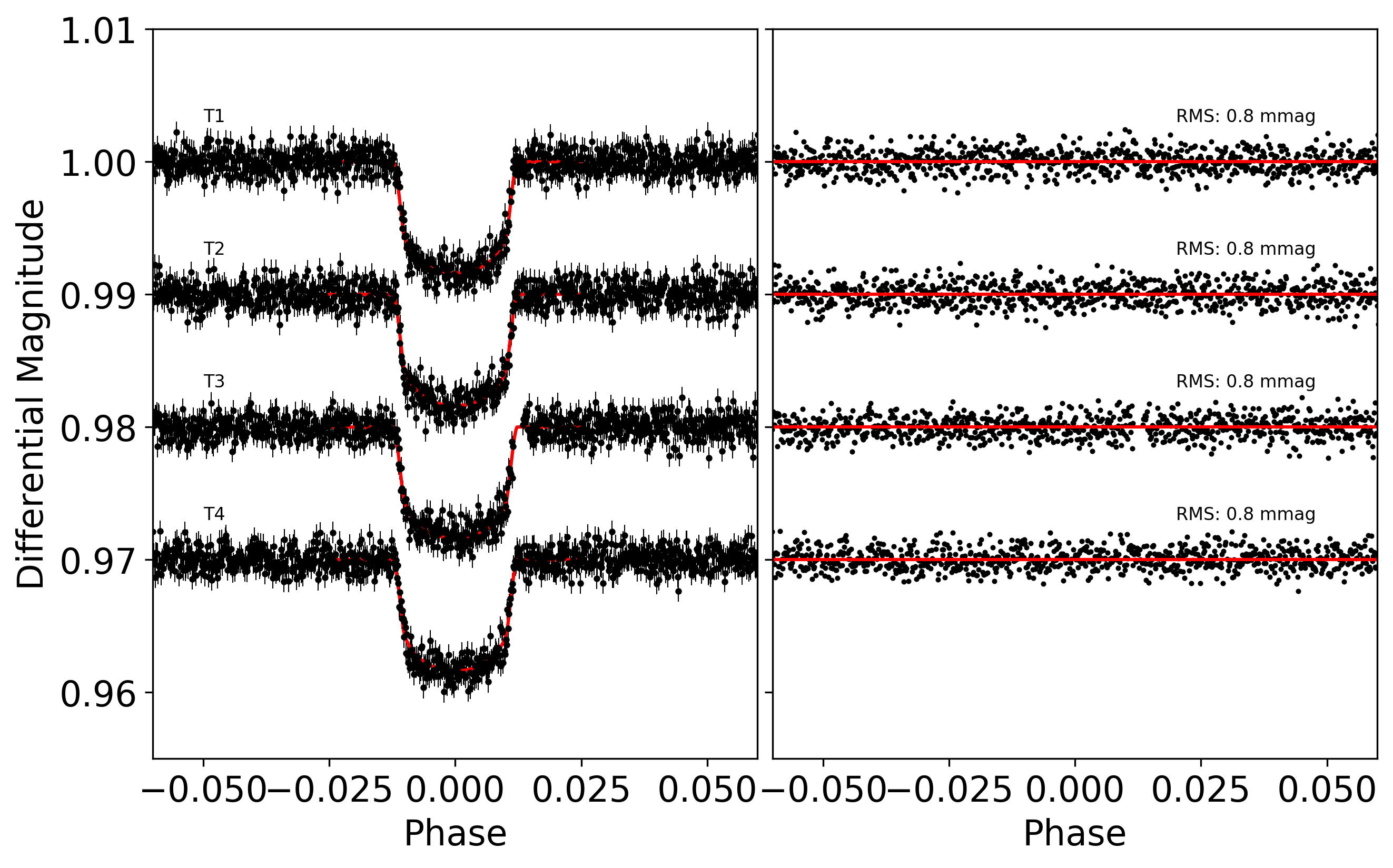}
    \includegraphics[width = \columnwidth]{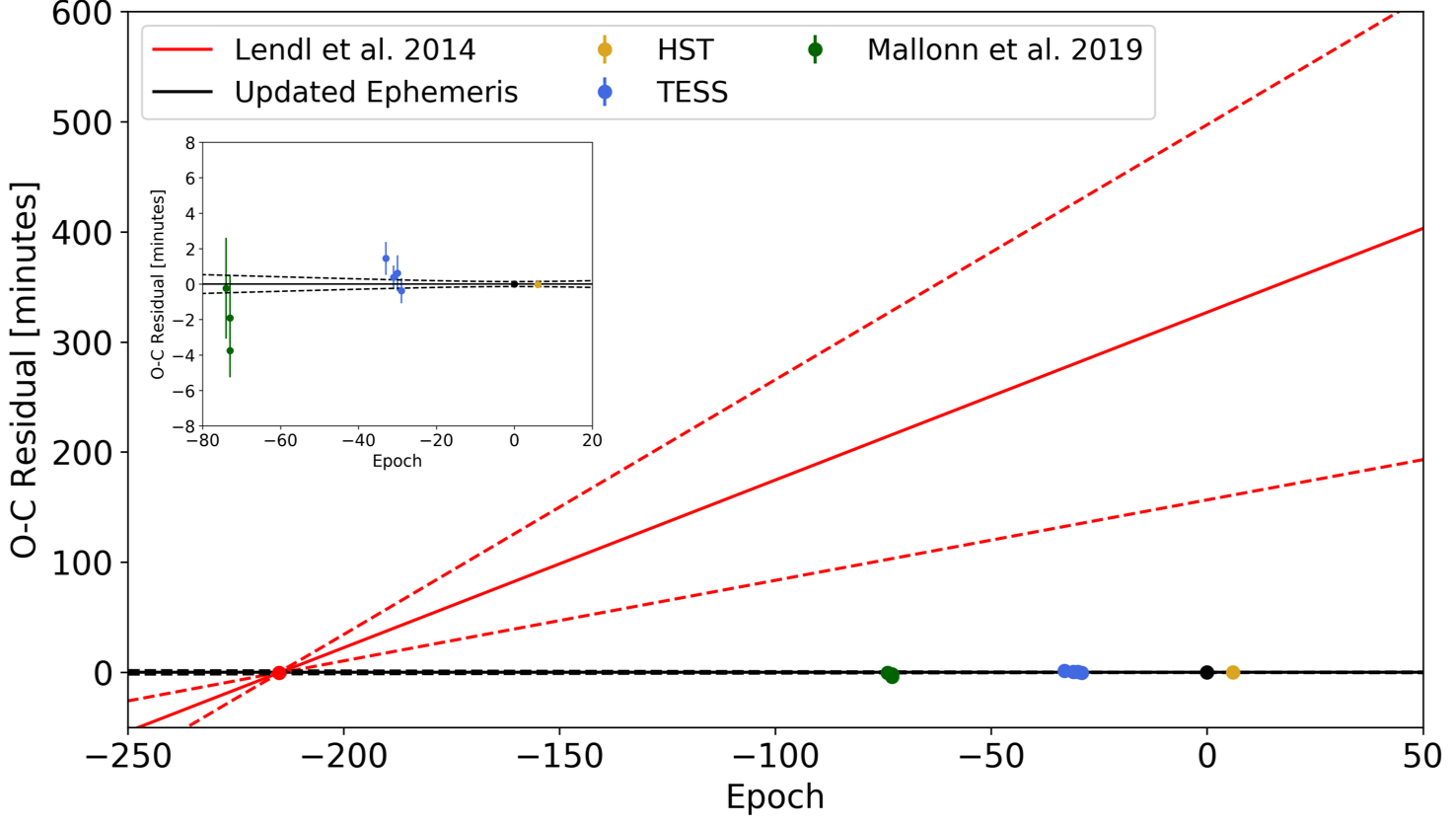}
    \caption{Top: TESS observations of WASP-117\,b presented in this work. Left: detrended data and best-fit model. Right: residuals from fitting. Bottom: Observed minus calculated (O-C) mid-transit times for WASP-117\,b. Transit mid time measurements from this work are shown in gold (HST) and blue (TESS), while the T$_0$ value for \citet{lendl_wasp117} is in red and the observations from \citet{mallonn} are in green. The black line denotes the new ephemeris of this work with the dashed lines showing the associated 1$\sigma$ uncertainties and the black data point indicating the updated T$_0$. For comparison, the previous literature ephemeris and their 1$\sigma$ uncertainties are given in red. The inset figure shows a zoomed plot which highlights the precision of the TESS and HST mid time fits.}
    \label{fig:k7_ephm}
\end{figure}

Our derived period is 1.2 s shorter than that from \citet{mallonn}. We improved the accuracy of the period and thus reduced the current uncertainty on the transit time with respect to the results from \citet{mallonn}. The observed minus calculated residuals, along with the fitted TESS light-curves are shown in Figure \ref{fig:k7_ephm} while the fitted mid times can be found in Table \ref{tab:mid_times}. Our new observations have been uploaded to ExoClock\footnote{\url{https://www.exoclock.space}}, a coordinated follow-up programme to keep transit times  up-to-date for the ESA Ariel mission.

\begin{table}
\centering
\begin{tabular}{cccc}
\hline\hline
\multicolumn{4}{c}{Retrieval Analysis Parameters}\\ \hline
Parameters & Prior bounds & Scale & Retrieved Value\\
\hline 
$R_p$ [$R_{J}$]& [0.8, 2] & linear  &  0.96$^{+0.02}_{-0.02}$
\\

$T_{term}$ [K]& [600, 1300] & linear  & 833$^{+260}_{-156}$
\\

$V_{\text{H}_2\text{O}}$ & [-12, -1] & $\log_{10}$  & -3.82 $^{+1.37}_{-1.55}$
\\

$V_{\text{CH}_4}$ & [-12, -1] & $\log_{10}$  &  unconstrained 
\\

$V_{\text{CO}}$ & [-12, -1] & $\log_{10}$ & unconstrained
\\

$V_{\text{CO}_2}$ & [-12, -1] & $\log_{10}$ & unconstrained
\\

$P_{cloud}$ & [6, -2] & $\log_{10}$  & 2.52 $^{+1.53}_{-1.25}$ \\
\hline\hline
\end{tabular}
\caption{List of the retrieved parameters, their uniform prior bounds, the scaling used and the corresponding retrieved posterior distribution mean values.}
\label{tab:retrieval_params}
\end{table}

\begin{table}
    \centering
    \begin{tabular}{ccc} \hline\hline
     Epoch & Transit Mid Time [BJD$_{TDB}$] & Reference  \\\hline\hline
  -216 & 2456533.824040 $\pm$ 0.000950 & \citet{lendl_wasp117} \\
  -75 & 2457946.728100 $\pm$ 0.001980 & \citet{mallonn} \\
  -74 & 2457956.749850 $\pm$ 0.001630 & \citet{mallonn} \\
  -74 & 2457956.751130 $\pm$ 0.001050 & \citet{mallonn} \\
  -34 & 2458357.571170 $\pm$ 0.000599 & This Work$^\dagger$ \\
  -32 & 2458377.613147 $\pm$ 0.000456 & This Work$^\dagger$ \\
  -31 & 2458387.633714 $\pm$ 0.000629 & This Work$^\dagger$ \\
  -30 & 2458397.654772 $\pm$ 0.000588 & This Work$^\dagger$ \\
  5 & 2458748.375370 $\pm$ 0.000082 & This Work* \\ \hline\hline
  \multicolumn{3}{c}{$^\dagger$Data from TESS *Data from Hubble}\\\hline \hline
    \end{tabular}
    \caption{Transit mid times used to refine the ephemeris of planets from this study.}
    \label{tab:mid_times}
\end{table}

{}

\subsection{Ariel \& JWST simulations}
\begin{figure}
    \centering
    \includegraphics[width=0.5\textwidth]{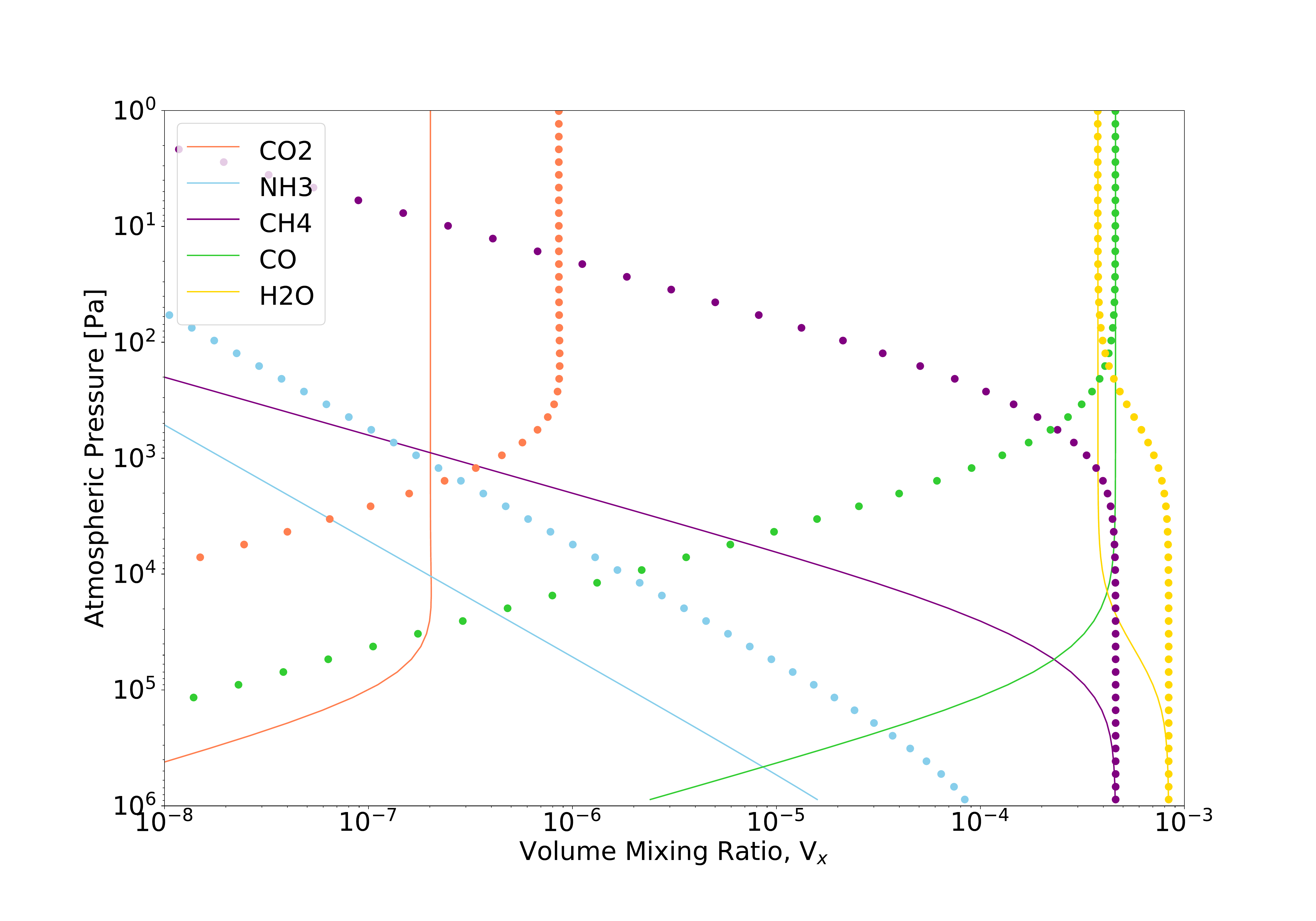}
    \caption{Equilibrium molecular abundance profiles as a function of atmospheric pressure for two different simulations of WASP-117\,b's atmosphere: dotted lines indicate cooler chemistry (800 K abundances); solid lines display hotter chemistry (1100 K abundances).}
    \label{fig:chem_abund}
\end{figure}

\begin{figure}
\centering 
 \includegraphics[width=0.5\textwidth]{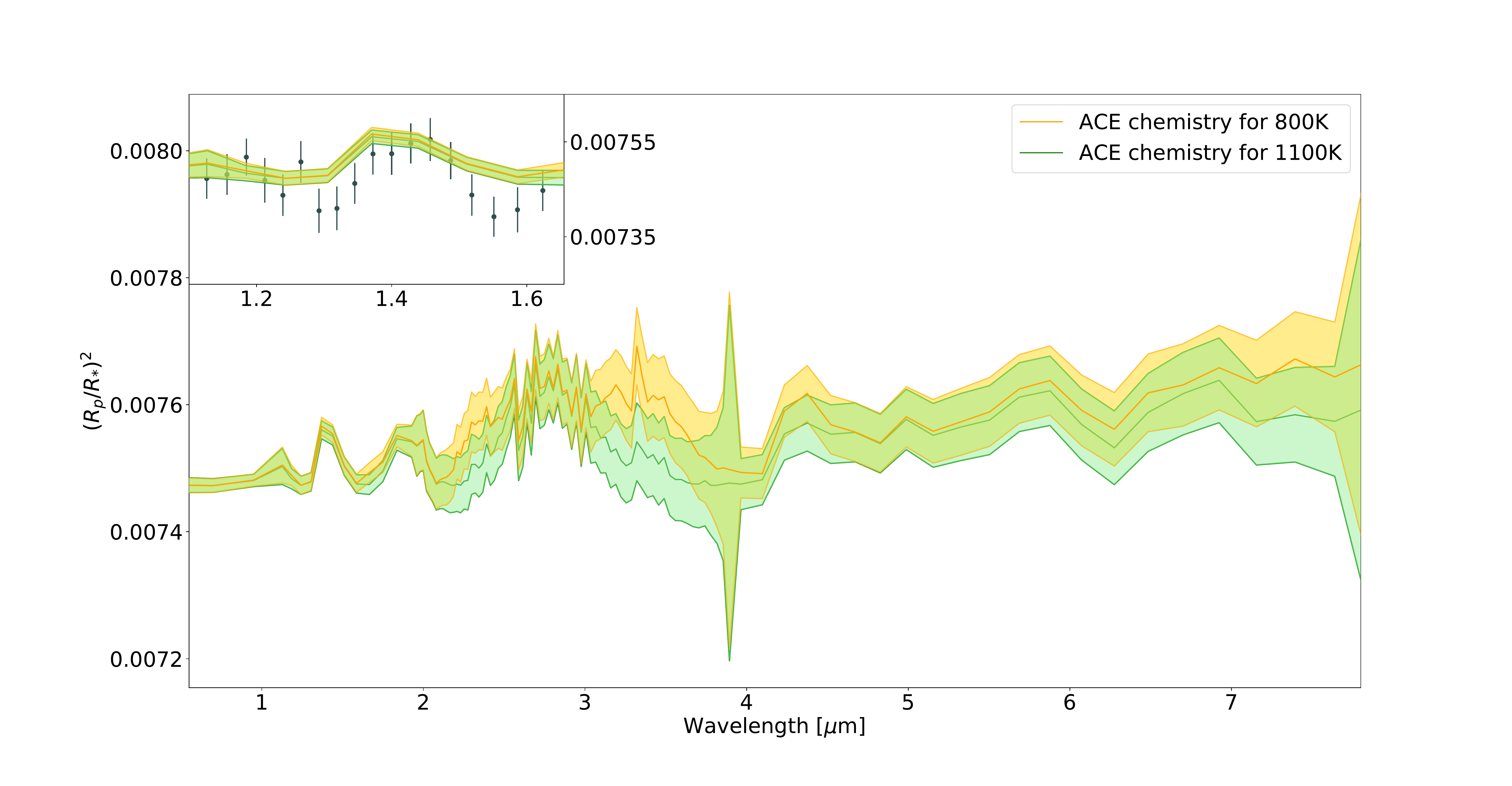}
 \includegraphics[width=0.5\textwidth]{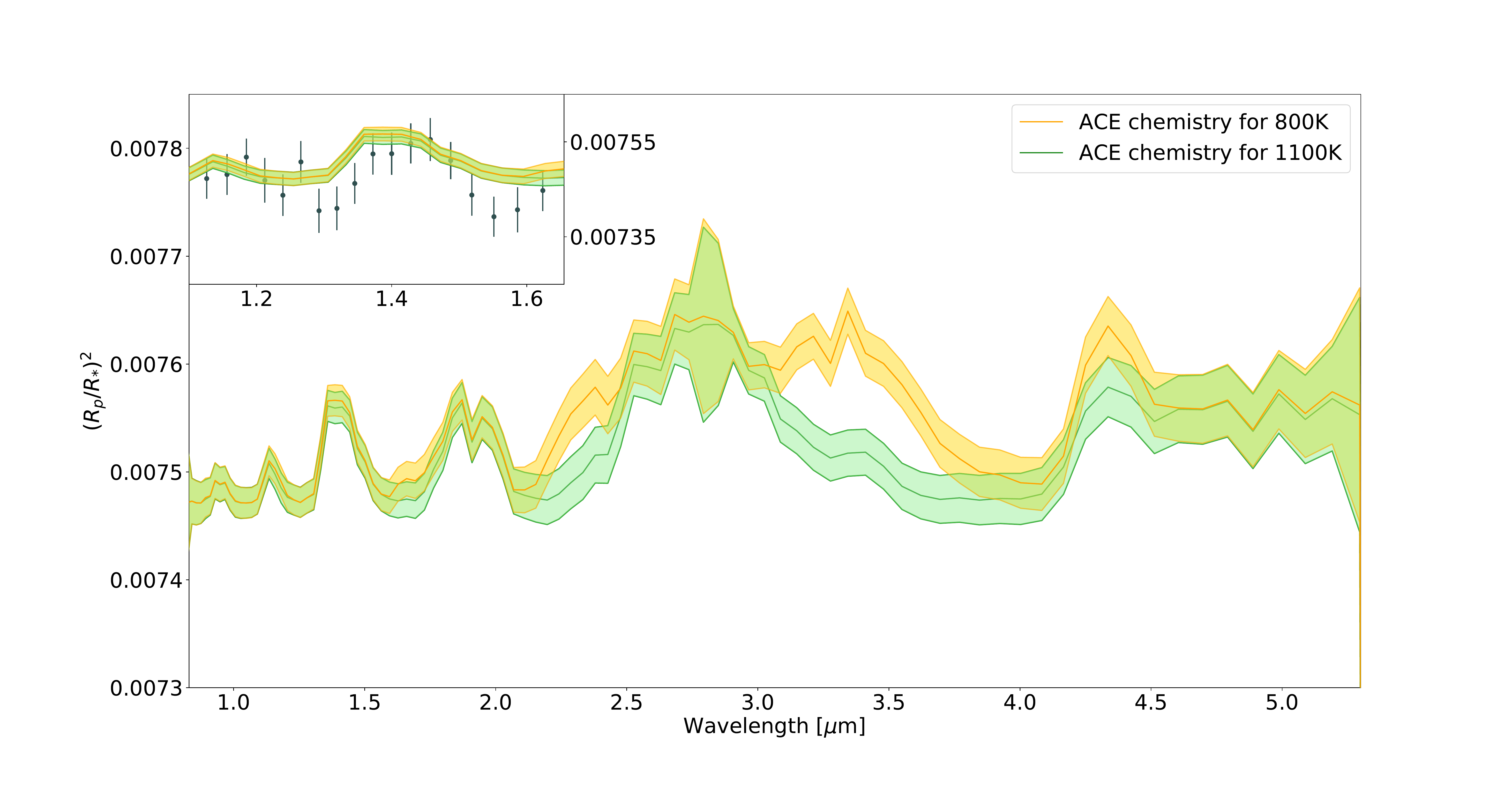}
 \caption{Simulated transmission spectra of WASP-117\,b, as observed by Ariel with 15 transits (top) and JWST with 4 transits (bottom). Forward models generated using chemical equilibrium abundances for 800 K are shown in yellow, whilst that for 1100 K are shown in green; with both spectra simulated using the retrieval temperature $T_{term} = 833$ K. Inset: zoom-in over the HST wavelength range with the observations over-plotted (not fitted).\vspace{5mm}} \label{fig:simulated}
 \end{figure}

The ACE equilibrium chemistry package identified H$_2$O, CH$_4$, CO, CO$_2$ and NH$_3$ as the most relevant species in the atmosphere of WASP-117\,b.
Resulting abundance profiles for these chemical species, as a function of atmospheric pressure, are displayed in Figure \ref{fig:chem_abund}, with the cooler-chemical atmosphere given by dashes, and the hotter-chemical atmosphere given by solid lines. The corresponding spectra created using these chemical profiles, but forward modelled using the retrieved terminator temperature of 833 K, as would be observed by Ariel and JWST, are displayed in Figure \ref{fig:simulated}. We can see that in the cooler chemistry scenario, there is a distinctly larger abundance of CH$_{4}$ in the region [$10^5, 10^2]$ Pa, the observable region, compared with the hotter chemical regime.

\begin{figure*}
    \centering
    \includegraphics[width=0.98\textwidth]{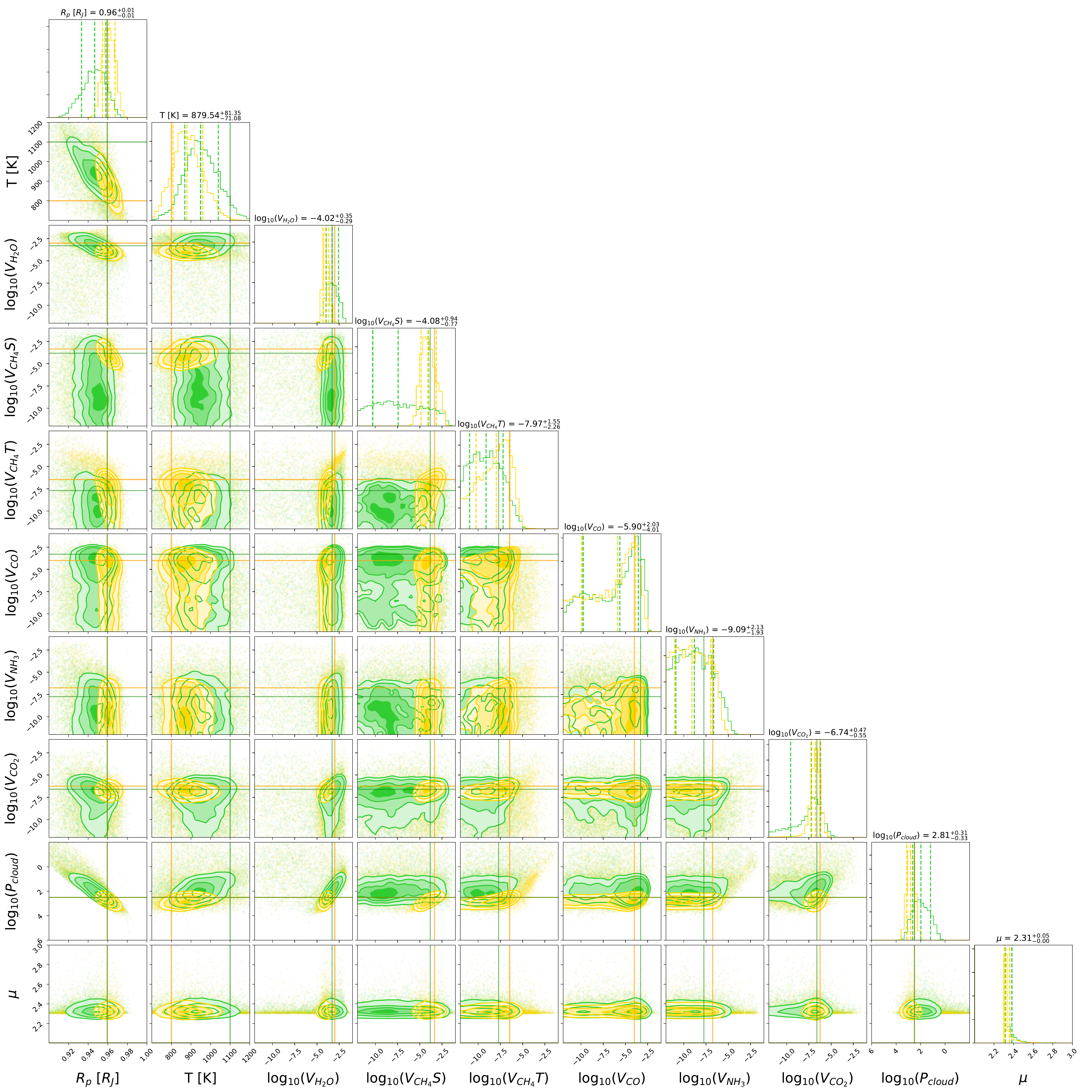}
    \caption{Posterior distributions over-plotted for a 2-layer retrieval of the simulated Ariel spectra given in Figure 8, with input from the 800 K and 1100 K atmospheres displayed in yellow and green, respectively. The posterior mean values presented correspond to the cooler atmospheric regime.}
    \label{fig:posteriorAriel}
\end{figure*}

\begin{figure*}
    \centering
    \includegraphics[width=0.98\textwidth]{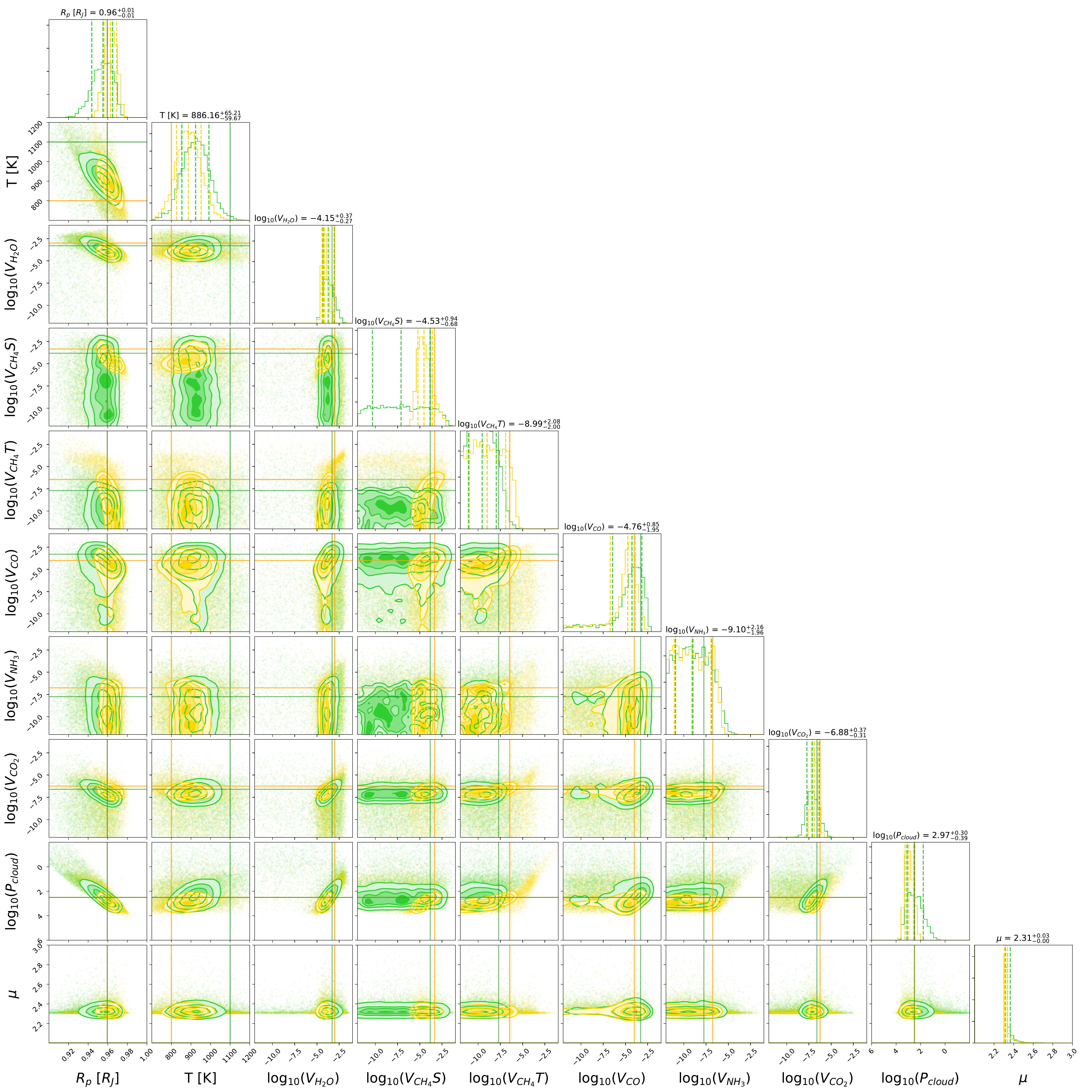}
    \caption{Posterior distributions over-plotted for a 2-layer retrieval of the simulated JWST spectra given in Figure 8, with input from the 800 K and 1100 K atmospheres displayed in yellow and green, respectively. The posterior mean values presented correspond to the cooler atmospheric regime.}
    \label{fig:posteriorJwst}
\end{figure*}

\vspace{3mm}
 In order to further constrain the CH$_4$ abundances, retrieval analysis was carried out for these two sets of simulated spectra, akin to that described in Section 2.3. We used prior distributions as specified in Table \ref{tab:retrieval_params}, with the exception of a two-layer profile for CH$_4$, with a pressure threshold between the two chemical profile layers set to $10^2$ Pa for the 800 K atmospheres, and to $10^{3.5}$ Pa for the 1100 K ones. The retrieved methane abundances at the ``surface'', at an atmospheric pressure larger than this threshold, and at the ``top'', at pressures smaller than this are denoted as $\log_{10}(V_{\text{CH}_{4}})_S$ and $\log_{10}(V_{\text{CH}_{4}})_T$, respectively. The resulting posterior distributions for the parameters of both nominal atmospheres are over-plotted in Figures \ref{fig:posteriorAriel} and \ref{fig:posteriorJwst} for the Ariel and JWST spectra, respectively. Over-plotted onto the posterior graphs are ``input'' chemical abundance values; for the species where constant abundances have been assumed the value has been extracted from Figure \ref{fig:chem_abund} at $10^3$ Pa, whilst the values for the methane abundances in the surface and top layers are taken at pressure points one order of magnitude either side of the pressure inflection points, for both sets of atmospheres. Both \citet{Changeat_2019} and \citet{changeat2020alfnoor} illustrate that the use of a 2-layer parameterisation, whilst significantly more revealing than an constant abundance profile, will only retrieve an abundance profile that aligns with the forward model at the peak of the molecular contribution function. Thus, our over-plotted abundance values are expected to be more accurate close to the pressure inflection points, but should be treated with caution at pressures elsewhere. Additionally, we note also that for the 1100 K atmospheres, molecular abundances prove harder to constrain but that forcing constant retrieval profiles on varied input abundance results in a bias on the retrieved temperature. This result is consistent with evidence for a retrieval bias towards lower determined temperatures than planetary effective temperatures, as investigated in \citet{macdonald2020cold}, \citet{Caldas_2019} and \citet{skaf_ares}.\\

Despite not being able to distinguish between the two regimes over the HST wavelength range, the distinction can in fact be made with 15 Ariel observations due to the CH$_4$ spectral features present due to ro-vibrational transitions at $2.3$, $3.3$ and $7.66 \mu m$ \citep{Yurchenko_2014}. As for JWST, only 4 transits (2 with NIRISS and 2 with NIRSpec G395M) are needed to illuminate this distinction. This makes WASP-117\,b a very promising candidate for observations with both Ariel and JWST, as with its bright host star and posited chemistry, this Saturn-mass planet could further illuminate exoplanet atmospheric chemical dynamics. 

\section{Discussion}
\subsection{Retrieval Results and Atmospheric Temperature}
Whilst the ADI of 2.30 provides positive but not strong evidence against the flat-line model, we recognise that this value is sensitive to the scattered region of the spectrum below $1.3\mu m.$ When the data points at 1.27$\mu m$ and 1.18$\mu m$ were removed and the same retrieval analysis performed, we obtain an ADI value of 4.30, moving over the threshold into strong evidence. The addition of WASP-117\,b to the long list of gaseous planets with prominent water features \citep{tsiaras_30planets,pinhas}, solidifies further the evidence that water appears to be ubiquitous in the atmospheres of such planets, with the presence of clouds in fact obscuring water that sits deeper in the atmosphere, and so weakening the observed transmission spectral signal \citep{sing}. 
\\

The terminator temperature was estimated to be  $T_{term}=833^{+260}_{-156}$ K, which sits in the range [677 K, 1093 K] at the 1$\sigma$ level. Our retrieval analysis uses a temperature prior informed by equilibrium temperature arguments as outlined in Section 2.3. Although atmospheric parameters such as geometric albedo and heat redistribution are not well constrained for this planet, what is significant from a chemical perspective is the possible fluctuation of temperature around $\approx 900$ K between apastron and periastron, since this is the threshold for which CO-CH$_4$ conversion oscillates between CO or CH$_4$ dominant in the $10^2-10^5$ Pa visible-region of the atmosphere, as is observed in \citet{Visscher_2012} for solar-metallicity gas. We take $\beta=1$ and $A=0.3$ to derive an equilibrium temperature range of [816 K, 1116 K], whereas \citet{lendl_wasp117} take $\beta=1$ and $A=0$ finding [897 K, 1225 K]. Despite not agreeing exactly, both temperature ranges include 900 K.
\\

We note that our retrieval analysis displays a significant degeneracy between the cloud pressure and water abundance. Degeneracies of this nature are common when only using data from WFC3 G141 \citep[e.g.][]{tsiaras_30planets} and the addition of data spanning visible wavelengths has been shown to remove this \citep{pinhas}. As TESS has also studied the transit of WASP-117\,b it could provide additional information to constrain parameters. However, the transit depth recovered from the TESS data is extremely shallow, around 200 ppm lower than the WFC3 data-set. \citet{carone_w117} also found an anomalously low TESS transit depth.
\\

Offsets between instruments can be caused for a number of reasons: due to imperfect correction of instrument systematics; from the use of different orbital parameters or limb darkening coefficients during the light curve fitting; or from stellar variability or activity \citep[e.g.][]{stevenson_gj436,stevenson_wasp12,tsiaras_30planets,alexoudi_inc,yip_lc,bruno_spots,aresIII,murgas_offset,yip_w96}. Here, we fitted the data-sets with the same limb darkening laws and orbital parameters, ruling out that potential explanation. \citet{carone_w117} studied whether stellar activity or the transit light source effect could be causing this discrepancy and found, while some offset could be explained, the magnitude of the offset was too great for this alone to be the cause. Therefore, we anticipate the offset being due to imperfect correction of instrument systematics. In the white light curve (Figure \ref{fig:white}) we noted the presence of some non-Gaussian residuals. While the divide-by-white method means the spectral light curves display Gaussian residuals, the whole spectrum may be shifted due to the imperfect white light curve fit. Further data, for example with the WFC3 G102 grism or CHEOPS, could help resolve this issue.

\subsection{Atmospheric Chemistry}

In order to understand the capabilities of future missions like JWST and Ariel to successfully probe the atmospheric chemistry of WASP-117\,b it is paramount to consider the detection limits of various chemical species with respect to the resolution and wavelength coverage of their instruments. Our simulated atmospheres for WASP-117\,b which we subsequently retrieve on are generated at the Ariel tier 2 resolution. The posterior distributions given in Figures \ref{fig:posteriorAriel} show that with 15 transits it is possible to reliably constrain the abundances of H$_2$O and of the carbon-based molecules CH$_4$ and CO$_2$, which is consistent with results from \citet{changeat2020alfnoor} in which upper atmosphere observational detection limits for tier 2 are determined to be $V_{x} \geq 10^{-7}$. For species like CO and NH$_3$, the detection limits are $V_{x} \geq 10^{-4}$, $10^{-7}$, respectively. Thus, in agreement with our simulations, it is clear that for Ariel, tracing these species will remain challenging. Correspondingly, simulated JWST spectra for 4 transits enable tighter constraints to be made upon retrieving, as illustrated in Figure \ref{fig:posteriorJwst}, but giving the same overall conclusions on detectability. In order to unify the power of both missions, but to avoid the well-known issues related to combining data-sets from different instruments as described in Section 4.1, results from each instrument should be used as prior knowledge to inform analysis with the other. For example, since the wavelength coverage of Ariel reaches as far into the visible as 0.5 $\mu$m, and further than JWST at 0.6 $\mu$m, these few additional data points could aid the removal of the degeneracy between retrieved cloud layer pressure and water vapour abundance. Additionally, as JWST has a greater sensitivity across the 3.0-3.5 $\mu m$ region, where the models are most distinctly separated, the NIRSpec G395M observations could be complementary to the Ariel data. As we have discussed, eccentric orbits cause a variation of atmospheric temperature due to variations in levels of stellar flux received by the planet as it traverses its orbit. As a result we can expect atmospheric chemical profiles to vary with time. A schematic diagram for the orbit of WASP-117\,b's orbit is displayed in Figure \ref{fig:orbit}, with the line-of-sight direction for which we observe the system during transit highlighted in yellow. In order to assess the chemistry that is observable during transit, chemical and dynamical mixing timescales must be considered with respect to the planet's orbital and spin periods. \citet{Visscher_2012} asserts that eccentric planet atmospheric chemistry is only affected by vertical mixing if the time elapsed between apastron and periastron exceeds vertical mixing timescales ($\tau_{mix} < \frac{1}{2}P_{orbital}$). Thus if, for now, we make the assumption that vertical mixing timescales are slower than this threshold, we can compare chemical timescales $\tau_{chem}(x)$ for a given molecular species $x$ to orbital timescales to assess what sort of chemistry might be observable during transit. 
 \\
 
 These chemical timescales are functions of pressure and temperature and so will oscillate throughout orbit; allowing perhaps for warmer chemistry to remain visible as the planet enters transit, despite the proximity of its point of ingress to apastron. As the temperature reaches 838 K at transit, the time taken to reach chemical equilibrium corresponding to this new atmospheric temperature may be larger than the time taken to travel there from periastron and thus the chemistry may not be able to adapt to the changes in the irradiation environment. \citet{Visscher_2012} calculates chemical timescales for various species in the atmospheres of the eccentric gaseous giant exoplanets HAT-P-2\,b and CoRoT-10\,b. In particular, they obtain $\tau_{chem}(\text{CH}_4) \approx \tau_{chem}(\text{CO}) \approx 10^{20}$ s in the observable region of $10^2 - 10^5$ Pa, with respect to CH$_4$-CO inter-conversion, at WASP-117\,b’s equilibrium periastron temperature of 838 K, which is indeed larger than half the planet’s orbital period of around $10^5$ s. 
 \\
 
 However, so far this argument only considers chemical equilibrium processes, which is almost certainly a gross simplification. In order to constrain chemical timescales more rigorously, a more in-depth analysis of the atmosphere which includes photo-dissociation processes, disequilibrium chemistry and vertical and horizontal mixing processes is required. However, in order to carry out preliminary analysis, we neglect vertical mixing, assume either tidal locking or a spin-synchronous orbit and consider only equilibrium chemical processes as a first-order approximation to assess feasibility of detectability. Under these assumptions we are left with two extremes for observable chemistry during transit. Assuming thermo-chemical equilibrium is achieved at periastron we may observe warmer periastron (1100 K) chemistry if the chemical timescales are slow enough such that the chemistry does not have time to adapt to the decreased temperature by the time it enters transit close to apastron ($\tau_{chem} > \frac{1}{2}P_{orbital}$). Or, on the contrary, if chemical timescales are fast enough to allow for the chemistry to adapt to the apastron temperature close to transit ($\tau_{chem} < \frac{1}{2}P_{orbital}$), or if chemical timescales are slow but chemical equilibrium is reached at apastron, we may observe cooler (800 K) apastron chemistry, i.e. what we expect. 
\\

A further caveat to this is that in the likely case that the planet spins, spin-orbit resonance may be able to somewhat homogenise the warmer and cooler regimes ($\tau_{mix} > P_{spin}$) possibly giving the planet a globally averaged chemistry between the hotter and cooler cases. In addition to the aforementioned chemical assumptions, other caveats to such a simplified analysis include the lack of a 3D atmospheric model, with self-consistent dynamics. \citet{Caldas_2019} illustrate that 3D analysis could enable the capture of the contamination of the terminator by the day-side of the planet, due to strong irradiation at the day-side, whilst \citet{macdonald2020cold} show that 1D retrieval analyses cannot capture compositional differences between the morning and evening terminator, whilst both effects could result in chemical inhomogeneities in the terminator region. A comprehensive study of all such processes would enable accurate characterisation of WASP-117\,b’s atmospheric chemistry, with the presented analysis of the two extremes of equilibrium chemistry serving to motivate follow-up study and observation.

\section{Conclusion}
WASP-117\,b's possession of an eccentric and misaligned orbit around its bright and stable F9 host makes it a tantalising object for atmospheric characterisation, since chemical timescales for CH$_4$ compete with the orbital period. With HST WFC3 observations, we present a retrieval solution with a well-constrained water vapour volume mixing ratio of $\log_{10}(V_{\text{H}_{2}\text{O}})= -3.82^{+1.37}_{-1.55}$, alongside a layer of opaque cloud. Carbon-based molecules such as CO, CO$_2$ and CH$_4$ prove harder to constrain using such data as the wavelength coverage and the signal-to-noise remains limiting. With the future telescopes JWST and Ariel, we present simulated spectra for WASP-117\,b as would be observed by these missions and show that it should be possible to probe possible variations in CH$_4$ chemistry in the observable region of the atmosphere.

\section*{Acknowledgements}

\textit{Data:} This work is based upon observations with the NASA/ESA Hubble Space Telescope, obtained at the Space Telescope Science Institute (STScI) operated by AURA, Inc. The publicly available HST observations presented here were taken as part of proposal 15301, led by Ludmila Carone \citep{carone_proposal}. These were obtained from the Hubble Archive which is part of the Mikulski Archive for Space Telescopes. This paper also includes data collected by the TESS mission, which are publicly available from the Mikulski Archive for Space Telescopes (MAST) and produced by the Science Processing Operations Center (SPOC) at NASA Ames Research Center \citep{jenkins_tess}. This research effort made use of systematic error-corrected (PDC-SAP) photometry \citep{smith_pdc,stumpe_pdc1,stumpe_pdc2}. Funding for the TESS mission is provided by NASA’s Science Mission directorate. We are thankful to those who operate this archive, the public nature of which increases scientific productivity and accessibility \citep{peek2019}.\\

\textit{Funding:} We acknowledge funding through the ERC Consolidator grant ExoAI (GA 758892) and the STFC grants ST/P000282/1, ST/P002153/1, ST/S002634/1 and ST/T001836/1. O.V. acknowledges the CNRS/INSU Programme National de Plan\'etologie (PNP) and CNES for funding support.\\

\software{Iraclis \citep{tsiaras_hd209}, TauREx3 \citep{al-refaie_taurex3}, pylightcurve \citep{tsiaras_plc}, ExoTETHyS \citep{morello_exotethys}, ArielRad \citep{mugnai}, ExoWebb \citep{exowebb}, Astropy \citep{astropy}, h5py \citep{hdf5_collette}, emcee \citep{emcee}, Matplotlib \citep{Hunter_matplotlib}, Multinest \citep{Feroz_multinest}, Pandas \citep{mckinney_pandas}, Numpy \citep{oliphant_numpy}, SciPy \citep{scipy}, corner \citep{corner}.}

\bibliographystyle{aasjournal}
\bibliography{main}

\end{document}